\documentclass[twocolumn,showpacs,preprintnumbers,amsmath,amssymb]{revtex4}
\usepackage{graphicx}
\usepackage{dcolumn}
\usepackage{bm}

\begin{document}


\title{High-sensitivity optical monitoring  of a micro-mechanical resonator \\
with a quantum-limited optomechanical sensor}

\author{O. Arcizet}
\author{P.-F. Cohadon}
\author{T. Briant}
\author{M. Pinard}
\author{A. Heidmann}
\affiliation{Laboratoire Kastler Brossel, Case 74, 4 place
Jussieu, F75252 Paris Cedex 05, France}
\thanks{Unit\'{e} mixte de recherche du Centre National de la Recherche
Scientifique, de l'Ecole Normale Sup\'{e}rieure et de
l'Universit\'{e} Pierre et Marie Curie}
\homepage{www.spectro.jussieu.fr/Mesure}

\author{J.-M. Mackowski}
\author{C. Michel}
\author{L. Pinard}
\affiliation{Laboratoire des Mat\'{e}riaux avanc\'{e}s, B\^atiment
VIRGO, Universit\'{e} Claude Bernard, 22 Bd Niels Bohr, F69622
Villeurbanne Cedex, France}

\author{O. Fran\c{c}ais}
\author{L. Rousseau}
\affiliation{Groupe ESIEE, ESYCOM Lab, GIS Micro Nano technologie,
Cit\'{e} Descartes BP 99, 2 Bd Blaise Pascal, F93162 Noisy le
Grand cedex, France}

\pagestyle{plain}

\begin{abstract}
We experimentally demonstrate the high-sensitivity optical
monitoring of a micro-mechanical resonator and its cooling by
active control. Coating a low-loss mirror upon the resonator, we
have built an optomechanical sensor based on a very high-finesse
cavity ($30\,000$). We have measured the thermal noise of the
resonator with a quantum-limited sensitivity at the
$10^{-19}\,\mathrm{m}/\sqrt{\mathrm{Hz}}$ level, and cooled the
resonator down to $5\,\mathrm{K}$ by a cold-damping technique.
Applications of our setup range from quantum optics experiments to
the experimental demonstration of the quantum ground state of a
macroscopic mechanical resonator.
\end{abstract}

\pacs{42.50.Lc, 05.40.Jc, 03.65.Ta}

\maketitle

{\it Introduction -} Optomechanical coupling between a moving
mirror and quantum fluctuations of light first appeared in the
context of interferometric gravitational-wave detection \cite
{Bradaschia90,Abramovici92} with the existence of the so-called
Standard Quantum Limit \cite{Caves81,Jaekel90,Braginsky92}. Since
then, several schemes involving a cavity with a movable mirror
subject to radiation pressure have been proposed either to create
non-classical states of both light \cite {Fabre94,Mancini94} and
mirror motion \cite {Bose97}, to perform Quantum Non Demolition
measurements \cite{Heidmann97}, or to entangle two movable mirrors
\cite{EPLEntanglement}. Recent progress in low-noise laser sources
and low-loss mirrors has made the field experimentally accessible
and has enlightened the unique sensitivity of interferometry, with
conventional fused silica mirrors \cite{EurophysLett,LaserCool} or
specially designed sensors such as a silicon torsion oscillator
\cite{Tittonen} or mirror-flexure system \cite{Australiens}.

Micro- and nano-electromechanical resonators play a great role in
the quest to detect quantum fluctuations of a mechanical resonator
\cite{Roukes,Cleland,Schwab} or for sensing purposes
\cite{Rugar,ClelandRoukes,MassSensing}. Detecting zero-point
motion of a mechanical oscillator requires high resonance
frequencies (up to the GHz band) and low temperature operation (in
the mK regime). It also requires a sufficient sensitivity on the
displacement measurement, which has not been reached yet by any
setup based upon an electrical detection scheme.

The optomechanical monitoring of a micro-mechanical resonator
therefore seems promising for the experimental observation of
quantum effects of radiation pressure and to reach the quantum
regime of a macroscopic oscillator \cite{SpieAustin}. The drastic
improvement of sensitivity is made at the expense of a larger
resonator and a correspondingly lower critical temperature, thus
requiring an active cooling strategy such as cold damping
\cite{LaserCool} or cavity cooling \cite{Karrai}.

In this Letter we demonstrate the experimental feasibility of
these concepts. We present an experiment where the motion of a
silicon micro-mechanical resonator is probed with an unprecedented
sensitivity by an optical setup based on a very high-finesse
optical cavity. We have observed the thermal noise spectrum of the
resonator over a wide bandwidth, and we have cooled the resonator
by a cold-damping technique. Quantum optics experiments possible
with such a setup are discussed.

\begin{figure}[b]
\begin{center}
\begin{tabular}{c}
\includegraphics[width=6 cm]{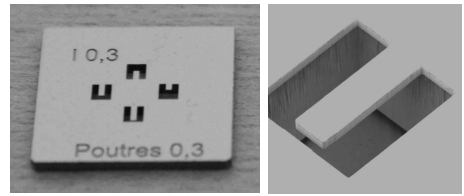}
\end{tabular}
\end{center}
\caption[example]{Left: optical image of a chip with 4
micro-resonators. Right: SEM image of one cantilever resonator.}
\label{fig:Mems}
\end{figure}

{\it Micro-resonator design and fabrication -} We have developed
silicon micro-mechanical resonators with typical transverse
dimensions from $400\,\mu \mathrm{m}$ to $1\,\mathrm{mm}$ and a
thickness of a few tens of microns. With such dimensions,
resonance frequencies are in the MHz range and the corresponding
effective masses down to the $\mu \mathrm{g}$ level. Fabrication
of the resonator proceeds as follows: we use
$1\,\mathrm{cm}$-squared chips, cut in a 4-inch SOI wafer
($60\,\mu\mathrm{m}$ Si$||2\,\mu \mathrm{m}\,SiO_2||500\,\mu
\mathrm{m}$ Si), each with up to 4 micro-resonator structures. The
structures are obtained by double-sided lithography and etched by
Deep Reactive Ion Etching \cite{DRIE}, which insures sharp edges.
We have fabricated resonators with different geometries, such as
the ones of Fig. \ref{fig:Mems}, but the results presented here
are all obtained with a $1\,\mathrm{mm}\times 1\,\mathrm{mm}\times
60\,\mu\mathrm{m}$ doubly-clamped beam. Each resonator chip is
coated on the upper side with a very high-reflectivity and
low-loss dielectric coating for 1064 nm.

\begin{figure}[t]
\begin{center}
\begin{tabular}{c}
\includegraphics[width=6 cm]{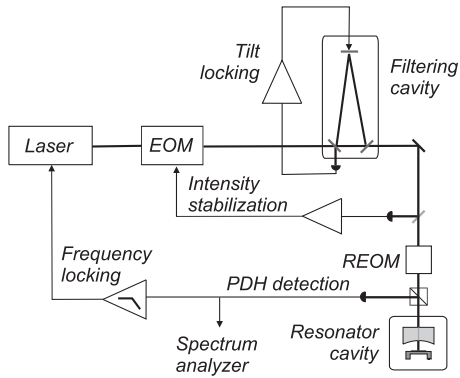}
\end{tabular}
\end{center}
\caption[example]{Experimental setup used to monitor the
displacements of the micro-mechanical oscillator. A Nd:YAG laser
is intensity-stabilized with an electro-optic modulator (EOM) and
spatially filtered before entering the resonator cavity. The
displacement signal is extracted by means of a Pound-Drever-Hall
phase modulation scheme using a resonant electro-optic modulator
(REOM). The low-frequency part of the signal is used to lock the
laser frequency to the cavity resonance.} \label{fig:SetupMesure}
\end{figure}

{\it Optomechanical sensing} - The optical monitoring setup is
based upon a single-ended Fabry-Perot cavity composed of the
micro-resonator as a totally reflecting back-mirror, and an input
coupling mirror with a $5\,\mathrm{cm}$ curvature radius (Fig.
\ref{fig:SetupMesure}). The cavity length is $2.4\,\mathrm{mm}$,
yielding an optical waist of $60\,\mu\mathrm{m}$. The resonator is
inserted in a mechanical structure which both guarantees optical
parallelism and allows for accurate translation of the resonator
perpendicularly to the optical axis in order to provide a fine
centering of the resonator. Due to the very good coating made on
the resonator, we have experimentally obtained a very high finesse
$\mathcal{F}= 30\,000$. From the input mirror transmission
$T=70\,\mathrm{ppm}$, this corresponds to overall losses (residual
transmission of the micro-resonator and losses of both mirrors)
equal to $L=140\,\mathrm{ppm}$.

The laser source is a highly-stabilized Nd:YaG laser operated at
$\lambda=1064\,\mathrm{nm}$. The laser beam is sent through a
wide-band electro-optic modulator (EOM) used as an
intensity-modulation device, and a triangular spatial filtering
cavity, locked onto resonance with the tilt-locking technique
\cite{TiltLocking}. Our laser source therefore delivers a perfect
$\mathrm{TEM}_{00}$ gaussian mode with well-defined intensity and
wavelength, and mode-matched to the high-finesse cavity by focussing
lenses.

In order to eliminate the drift between the laser and the optical
cavity, the cavity is temperature-stabilized with residual
temperature fluctuations below $10\,\mathrm{mK}$. The laser
frequency is finally locked at resonance by the Pound-Drever-Hall
(PDH) technique via a resonant electro-optical modulator (REOM)
which provides a phase modulation of the incident beam at the
sideband frequency of $12\,\mathrm{MHz}$. The resonator
displacements are monitored by the high-frequency part of the PDH
error signal, and the displacements are calibrated by comparison
with the effect of a frequency modulation of the laser beam
\cite{EurophysLett}.

\begin{figure}[t]
\begin{center}
\begin{tabular}{c}
\includegraphics[width=6.5 cm]{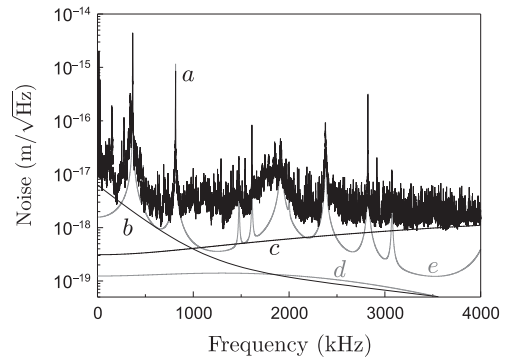}
\end{tabular}
\end{center}
\caption[example]{Noise amplitude of the Pound-Drever-Hall signal,
calibrated as micro-resonator displacements, over a
$4\,\mathrm{MHz}$ span ({\it a}). Other curves represent relevant
noises: maximum frequency noise level ({\it b}), shot-noise level
({\it c}), and optical index fluctuations due to the residual
pressure ({\it d}). Curve ({\it e}) is the thermal noise spectrum
expected from FEM simulations.} \label{fig:ScanLarge}
\end{figure}

{\it Noise levels and sensitivity -} Curve {\it a} of Fig.
\ref{fig:ScanLarge} presents the resulting calibrated noise
spectrum obtained at room temperature with a resolution bandwidth
of $20\,\mathrm{Hz}$ and for an incident laser intensity of
$1.5\,\mathrm{mW}$. The spectrum exhibits sharp peaks with high
dynamics, associated to the acoustic modes of the micro-resonator.

Curves {\it b} to {\it d} of Fig. \ref{fig:ScanLarge} represent
other relevant noises. The frequency noise of the laser has been
independently characterized with the filtering cavity. Curve {\it b}
presents an overestimated envelope. Frequency noise does not affect
the sensitivity of our experiment for frequencies higher than
$500\,\mathrm{kHz}$ and its effect could be further reduced by using
a shorter cavity. The optical cavity is operated in vacuum (with a
residual pressure below $10^{-2}\,\mathrm{mbar}$) in order to
minimize the effect of optical index fluctuations. Curve {\it d}
presents the expected noise level deduced from the measurement made
at ambient pressure, in agreement with a simple theoretical model
\cite{Japonais}.

At frequencies above $1\,\mathrm{MHz}$, the sensitivity is only
limited by the quantum phase noise of the reflected field (curve
{\it c}) to a level
\begin{equation}
\delta x_{\rm min}= \frac{\lambda}{16{\cal
F}\sqrt{I}}\frac{F(m)}{\sqrt{\eta\eta_{\mathrm{ph}}}}\frac{T+L}{T}
\sqrt{1+\left(\frac{f}{\Delta\nu}\right)^2},
\end{equation}
where $I$ is the mean incident intensity (counted as
$\mathrm{photons/s}$), $f$ the analysis frequency, $\Delta\nu =
1.05\,\mathrm{MHz}$ the cavity bandwidth, $\eta = 91\%$ the mode
matching of the beam to the cavity, $\eta_{\mathrm{ph}} = 93\%$
the detection efficiency and $F(m)$ a function of the modulation
index $m$ of the PDH scheme. With our parameters, the
quantum-limited sensitivity is equal to
$4\times10^{-19}\,\mathrm{m}/\sqrt{\mathrm{Hz}}$ at 1 MHz.

\begin{figure}[t]
\begin{center}
\includegraphics[width=6 cm]{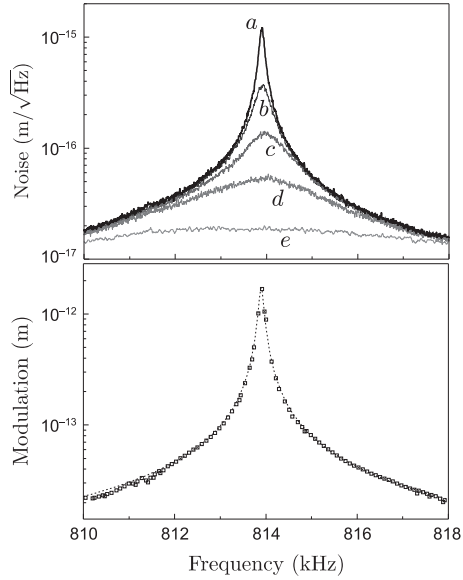}
\end{center}
\caption[example]{Top: thermal noise amplitude spectra around a
mechanical resonance of the oscillator at room temperature ({\it
a}), and at lower effective temperatures obtained by cold damping
({\it b} to {\it e}). Bottom: mechanical response to a modulated
electrostatic force (squares) and corresponding lorentzian fit
(dashed line).} \label{fig:Pic813}
\end{figure}

{\it Single mode optomechanical characterization -} The elasticity
theory and the fluctuation-dissipation theorem \cite{Saulson}
allows one to account for the observed thermal noise spectrum,
which can be seen as the sum of thermal peaks and off-resonance
tails of the vibration modes. Our setup allows to study every
vibration mode in great details. As an example, curve {\it a} of
Fig. \ref{fig:Pic813} presents the noise spectrum acquired over a
$4\,\mathrm{kHz}$ span centered around the mechanical resonance at
814 kHz. A lorentzian fit of the resonance gives access to the
optomechanical characteristics of the mode: resonance frequency
$f_{\mathrm{m}}\simeq814\,\mathrm{kHz}$, effective mass
$m_{\mathrm{eff}}\simeq 190\,\mu\mathrm{g}$, in good agreement
with the expected values ($890\,\mathrm{kHz}$ and
$130\,\mu\mathrm{g}$), computed with a finite element method
(FEM). The discrepancy decreases quickly and is below $5\%$ for
higher frequency modes. It appears to be mainly due to the
coupling of the resonator modes with the wafer modes, as shows the
dependence of the computed frequencies with the location of the
resonator over the chip. The mechanical quality factor $Q$ varies
from $5\,000$ to $15\,000$ among the modes, notably enhanced by
the vacuum operation.

An external force can be applied onto the resonator: we have used
an electrostatic force via a voltage modulation and an offset
applied between the resonator and a tip. The resulting force is on
the order of $1\,\mathrm{nN}$. The bottom curve of Fig.
\ref{fig:Pic813} shows the corresponding mechanical response and
confirms the mechanical origin of the resonance, along with the
values of the parameters
($f_{\mathrm{m}},\,m_{\mathrm{eff}},\,Q$).

\begin{figure}[t]
\begin{center}
\includegraphics[width=5.7 cm]{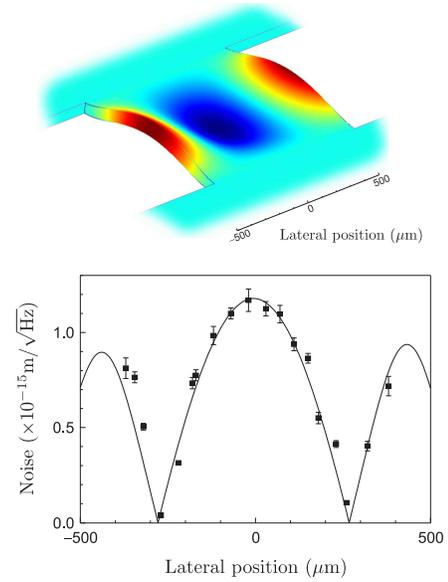}
\end{center}
\caption[example]{Top: computed spatial profile of the
$814\,\mathrm{kHz}$ vibration mode of the resonator. Bottom:
variations of the observed displacement noise level as a function
of the lateral position of the optical spot. Dots: Experimental
points; full line: fit with the expected spatial profile. The
vertical error bars are mainly due to the optical finesse
variation, especially at the edge of the resonator.}
\label{fig:EffMassSpot}
\end{figure}

{\it Spatial profiles -} As the observed displacements depend on
the overlap between the spatial structure of the mode and the
optical intensity profile inside the cavity \cite{ModesSpatial},
the spatial profile can be mapped by translating the resonator
with respect to the laser beam: Fig. \ref{fig:EffMassSpot}
presents the measured thermal noise level as a function of the
transverse displacement between the resonator and the laser waist.
The results are in excellent agreement with the noise levels
expected from the computed spatial structure of the mode. This
sheds new light onto the variations of the mechanical quality
factors observed between the various modes: one gets a low value
for a mode where the vibration evolves along the longitudinal
direction of the beam, whereas the value is higher for a
`transverse' mode with a low displacement at the clamping location
\cite{CrossLifshitz}.

{\it Multimode spectrum -} Performing an individual study of each
acoustic modes of the resonator allows us to quantitatively explain
the observed multimode noise spectrum of Fig. \ref{fig:ScanLarge}.
Curve {\it e} presents the expected thermal noise spectrum of the
resonator: only modes predicted by the FEM were taken into account,
with the experimental values
($f_{\mathrm{m}},\,m_{\mathrm{eff}},\,Q$) obtained by fitting the
individual thermal noise spectra.

The yet unmodelled discrepancy, for example around 1800 kHz,
appears to be due to the coupling between the resonator modes and
neighboring modes of the silicon chip. This can be further
accounted for by a noise spectrum monitored outside the resonator,
on the wafer surface, which clearly exhibits the same vibration
modes, along with a number of smaller peaks due to the modes of
the silica input coupling mirror, which also constitute the
quasi-continuum observed at higher frequencies.

{\it Cold damping -} We have finally demonstrated the possibility
to cool the resonator by a cold-damping feedback mechanism
\cite{LaserCool}. The monitoring signal is used in a feedback loop
to apply a controlled additional viscous force to the resonator,
without any additional noise. According to the
fluctuation-dissipation theorem, this yields a lower effective
temperature. Curves {\it b} to {\it e} of Fig. \ref{fig:Pic813}
present the thermal noise spectra of the resonator obtained for
increasing feedback gains. Due to the additional damping, the
lorentzian shapes are widened, whereas the amplitudes are strongly
reduced. The reduction of the curve area is directly related to
the effective temperature by the equipartition theorem. We have
reached a temperature of $5\,\mathrm{K}$, corresponding to a
cooling factor of 60.

{\it Conclusion} - We have presented an experiment where the
motion of a micro-mechanical resonator is monitored at the
$10^{-19}\,\mathrm{m}/\sqrt{\mathrm{Hz}}$ level with a setup based
upon a stabilized laser source and a very high-finesse optical
cavity. The motion and the optomechanical behavior have been fully
studied and accounted for, both at and off-resonance, at
frequencies of interest for quantum optics experiments ($\ge
500\,\mathrm{kHz}$). This is to our knowledge the first monitoring
of the motion  of a micro-mechanical resonator over such a large
frequency band. Our setup also presents a thousandfold-improvement
in sensitivity over any previous detection scheme used to monitor
the displacement of a micro-resonator, either electrical or
optical \cite{Cleland,Schwab,Ekinci,Hoogenboom}. However, there is
still room for improvement, both optical and mechanical: the
cavity finesse achieved so far is mainly limited by the roughness
of the commercial silicon wafer, and higher values have already
been obtained for the mechanical quality factor
\cite{RoukesDissipation}. The design of a resonator with smaller
size and with a lower impact of the wafer upon its motion is
currently under investigation.

Low temperature operation of such a resonator opens the way to
quantum optics experiments, as well as the experimental
observation of the quantum ground state of a macroscopic
mechanical resonator. Though our resonators operate in the MHz
range and therefore require a more stringent condition on the
temperature, the sensitivity achieved is promising and a specific
scheme, based upon active cooling by a cold damping mechanism
\cite{LaserCool} has already been proposed \cite{SpieAustin}. The
single-mode resonant behavior observed over more than 40 dB with
our resonator and the corresponding cooling by two orders of
magnitude we have obtained seem especially promising in that
purpose.

\smallskip
Acknowledgements are due to Francesco Marin and to Ping Koy Lam
for providing us respectively with the input mirror of our
measurement cavity and the mirrors of our filtering cavity, and to
Vincent Loriette for the characterization of our resonators.


\end{document}